\documentclass[lettersize, journal]{IEEEtran}
\usepackage{setspace}
\usepackage{romannum}
\usepackage{caption}
\usepackage{subfigure}
\usepackage{bm}
\usepackage{upgreek}
\usepackage{amsmath,amsfonts}
\usepackage{algorithmic}
\usepackage{algorithm}
\usepackage{array}
\usepackage{textcomp}
\usepackage{stfloats}
\usepackage{url}
\usepackage{verbatim}
\usepackage{graphicx}
\usepackage{cite}

\usepackage{tcolorbox}
\usepackage{enumitem}
\usepackage{xcolor}

\newtcolorbox{aibox}{
	colback=blue!10!white,
	colframe=blue!50!black,
	boxrule=0.5pt,
	arc=4pt,
	boxsep=0pt,
	left=6pt,
	right=6pt,
	top=6pt,
	bottom=6pt,
	before upper={\textbf{LLM:} }
}

\newtcolorbox{userbox}{
	colback=green!10!white,
	colframe=green!50!black,
	boxrule=0.5pt,
	arc=4pt,
	boxsep=0pt,
	left=6pt,
	right=6pt,
	top=6pt,
	bottom=6pt,
	before upper={\textbf{User:} }
}

\usepackage{fancyhdr}

\begin{document}
	
	\title{Large Language Model-Based Task Offloading and Resource Allocation for Digital Twin Edge Computing Networks}
	
	\author{Qiong Wu,~\IEEEmembership{Senior Member,~IEEE,} Yu Xie, Pingyi Fan,~\IEEEmembership{Senior Member,~IEEE,} \\Dong Qin, Kezhi Wang,~\IEEEmembership{Senior Member,~IEEE,} Nan Cheng,~\IEEEmembership{Senior Member,~IEEE,} \\and Khaled B. Letaief, ~\IEEEmembership{Fellow,~IEEE}
		% <-this % stops a space
		\thanks{This work was supported in part by Jiangxi Province Science and Technology Development Programme under Grant No. 20242BCC32016, in part by the National Natural Science Foundation of China under Grant No. 61701197, in part by the National Key Research and Development Program of China under Grant No. 2021YFA1000500(4), in part by the Research Grants Council under the Areas of Excellence Scheme under Grant AoE/E601/22R and in part by the 111 Project under Grant No. B23008.  }% <-this % stops a space
		\thanks{Qiong Wu and Yu Xie are with the School of Internet of Things Engineering, Jiangnan University, Wuxi 214122, China, and also with the School of Information Engineering, Jiangxi Provincial Key Laboratory of Advanced Signal Processing and Intelligent Communications, Nanchang University, Nanchang 330031, China (e-mail: qiongwu@jiangnan.edu.cn, yuxie@stu.jiangnan.edu.cn).}
		\thanks{Pingyi Fan is with the Department of Electronic Engineering, State Key
		laboratory of Space Network and Communications, Beijing National Research Center for Information Science and Technology, Tsinghua University, Beijing 100084, China (e-mail: fpy@tsinghua.edu.cn).}
		\thanks{Dong Qin is with the School of Information Engineering, Jiangxi Provincial Key Laboratory of Advanced Signal Processing and Intelligent Communications, Nanchang University, Nanchang 330031, China (email: qindong@seu.edu.cn).}
		\thanks{Kezhi Wang is with the Department of Computer Science, Brunel University, London, Middlesex UB8 3PH, U.K. (email: Kezhi.Wang@brunel.ac.uk).}
		\thanks{Nan Cheng is with the State Key Laboratory of ISN and the School of Telecommunications Engineering, Xidian University, Xi'an 710071, China  (e-mail: dr.nan.cheng@ieee.org).}
		\thanks{Khaled B. Letaief is with the Department of Electrical and Computer Engineering, the Hong Kong University of Science and Technology, Hong Kong (email: eekhaled@ust.hk).}}
	
	
	
	\maketitle
	
	\begin{abstract}
		In this paper, we propose a general digital twin edge computing network comprising multiple vehicles and a server. Each vehicle generates multiple computing tasks within a time slot, leading to queuing challenges when offloading tasks to the server. The study investigates task offloading strategies, queue stability, and resource allocation. Lyapunov optimization is employed to transform long-term constraints into tractable short-term decisions. To solve the resulting problem, an in-context learning approach based on large language model (LLM) is adopted, replacing the conventional multi-agent reinforcement learning (MARL) framework. Experimental results demonstrate that the LLM-based method achieves comparable or even superior performance to MARL.
		
	\end{abstract}
	
	\begin{IEEEkeywords}
		  Large language model, Digital twin, Resource allocation, Edge computing
	\end{IEEEkeywords}
	
	\section{Introduction}
	\IEEEPARstart{T}{he} growth of the Internet of things (IoT) has greatly expanded vehicular applications, improving the driving experience\cite{ref1, ref2, ref3}. However, this also leads to the generation of a large number of computing tasks. Vehicles' limited computing resources make independent task handling impractical \cite{ref4,ref5,ref6,ref7}. This limitation often results in significant delays due to the inability to process tasks promptly, thereby adversely affecting the quality of service (QoS) of the applications.
	
	To address this issue, vehicular edge computing (VEC) offers a promising solution. VEC enables vehicles to offload tasks to nearby servers by deploying VEC nodes or roadside units along the road. These servers leverage their abundant computing resources to process the diverse tasks generated by vehicles and return the results efficiently\cite{ref8, ref9, ref10, ref11}. However, VEC introduces new challenges. Environmental dynamics and vehicle mobility often create uncertainty in the volume of tasks offloaded by each vehicle\cite{ref12, ref13, ref14, ref15}. This uncertainty leads to competition among vehicles for the server’s computing resources, ultimately affecting the processing efficiency of offloaded tasks\cite{ref16, ref17, ref18, ref19}. As a result, effectively scheduling task offloading and allocating server resources becomes a critical challenge.
	
	Digital twin enables the simulation, analysis, and optimization of physical entities by creating digital replicas of real-world objects\cite{ref20, ref21, ref22, ref23}, such as vehicles, industrial machinery, and aviation engines. These digital replicas establish a full life cycle model for their corresponding physical entities\cite{ref24}. In addition, IoT advancements have greatly enhanced digital twin capabilities. They now support not only unidirectional information mirroring but also bidirectional interaction through intra-twin communication\cite{ref25}. Vehicles in dynamic environments can be virtually modeled by leveraging digital twin technology. Their digital replicas capture relevant vehicle information through intra-twin communication, enabling flexible task offloading and dynamic scheduling of server resources\cite{ref26,ref27,ref28}. In a digital twin edge computing network, a vehicle may run multiple applications simultaneously, generating diverse computational tasks within a single time slot, rather than a single task as previously assumed. Additionally, task offloading by multiple vehicles may impact edge server queue stability. Therefore, developing a task offloading and resource allocation scheme that ensures long-term queue stability becomes critically important. 
	
	The popular way to face the above problems is the deep reinforcement learning (DRL). The actions taken are gradually optimized through the constant interaction of the agent with the environment until the optimal action is obtained for the highest long-term reward. However, this approach is not without its drawbacks. DRL requires a lot of effort to adjust the model parameters in the process of training the model, and non-optimal parameters may cause the model to fall into a local optimum. LLM, as a novel approach, due to its strong capability resulting from multimodal pre-training, can save a lot of time and effort in the inference, and only need to design the prompts for LLM. Moreover, it is possible to further optimize the scheme obtained by MARL with the help of LLM method.
	
	In this paper, we construct a digital twin edge computing network with a base station and a server\footnote{The source code has been released at: https://github.com/qiongwu86/-LLM-Based-Task-Offloading-and-Resource-Allocation-for-DTECN}. Facing the situation that vehicles generate multiple types of computing tasks in a single time slot, we analyze the delay and energy consumption of each type of task under the influence of digital twin, and propose a in-context learning method based on LLM to optimize the average quality of service as well as the total energy consumption of the whole system scenario. The main contributions of this paper are as follows:
	
	\begin{itemize}
		\item We propose a generic digital twin edge network scenario and analyze the server queue backlog issue arising from vehicles generating heterogeneous tasks within a single time slot. Furthermore, we examine the subsequent impact of these queued tasks on both the system's average QoS and total energy consumption. To jointly optimize these two key metrics, we formulate and solve a corresponding optimization problem.
		%We construct a generic scenario of the digital twin edge network including multiple vehicles, a single base station, and a single server. We analyze the server queue backlog problem caused by vehicles generating multiple types of tasks within a single time slot in this scenario, and subsequently analyze the system average quality of service and total energy consumption affected by the various types of tasks after entering the queue. Based on the objective of optimizing the average QoS and total energy consumption, we construct a related optimization problem to solve it.
		\item We apply Lyapunov optimization to the constraints involving queue stability in the original problem, converting it from a long-term stability problem to a short-term decision problem, and consequently reconstructing the original problem.
		%To address the constraint of long-term queue stability in the optimization problem, the Lyapunov optimization method is employed. This approach transforms the original problem into a reconstructed optimization framework, converting it from a long-term stability challenge into a series of short-term decision-making problems.
		\item We propose a MARL-guided context-based LLM approach to the problem. Specifically, the LLM is made to output solutions by designing prompts for the LLM and combining them with a case set provided by MARL.
		%To solve the problem,  we use an in-context learning method based on LLM by designing and tuning the prompt input to LLM. Specifically, we construct the prompt as task goal, task definition, and additional rule, and we also provide a small number of cases for LLM to enhance model learning. We analyze the performance of the in-context learning algorithm by comparing it with the MARL algorithm.
	\end{itemize}
	
	\section{Related work}
	In this section, we first review related work on digital twins in vehicular edge network, and then survey existing studies on large models for connected car scenarios.
	
	\subsection{\textit{Digital Twin in VEC}}
	There has been some literature on the utilization of DT under VEC. In \cite{ref12}, Dai \textit{et al}. used DT as well as deep reinforcement learning (DRL) to optimize the offloading delay under adaptive DT network. In \cite{ref29}, Zhang \textit{et al}. utilized digital twins to guide vehicles to aggregate edge services to minimize offloading costs. In \cite{ref30}, Liao \textit{et al}. developed a DT model about a driver to predict his behavior in the context of an autonomous vehicle and a human-driven vehicle. In \cite{ref31}, Zhang \textit{et al}. constructed a vehicle edge caching system by DT with VEC and used DRL to formulate the optimal scheme. In \cite{ref32}, Zheng \textit{et al}. proposed a DT-based prediction model to predict the waiting time for a vehicle to connect to the network while using DRL to minimize the long-term delay and energy consumption. In \cite{ref33}, Zhao \textit{et al}. input the global information of vehicles into the DT model to assist the clustering algorithm in reducing the task offloading range for the purpose of offloading prediction. In \cite{ref34}, Li \textit{et al}. modeled DTs for roadside units(RSUs), UAVs, and vehicles to assist in task offloading for vehicles and management of resources for RSUs and UAVs. These literatures have used the DRL approach or machine learning algorithms in solving optimization problems, and no literature has yet used the large model approach to solve the problem.
	
	\subsection{\textit{Large Language Model}}
	At present, the research on LLM for Edge Computing is still in the exploratory stage. In \cite{ref35}, Zhou \textit{et al}. used LLM to investigate the case of power allocation for the base station, and the experiments show that LLM is able to avoid the tedious model training and hyper-parameter fine-tuning of machine learning, while achieving performance results comparable to those of traditional DRL. In \cite{ref36}, Zhou \textit{et al}. used generative AI for radio resource allocation and task offloading in an edge-cloud network to compute the content generation delay via LLM and optimize the offloading decision using LLM. Simulations show that the proposed context learning approach achieves satisfactory results without specialized model training and fine-tuning compared to traditional DRL. In \cite{ref37}, Lee \textit{et al}. developed an LLM-based resource allocation method for wireless communication systems to maximize energy and spectrum efficiency, and confirmed the applicability and feasibility of the scheme through experiments. In \cite{ref38}, Fu \textit{et al}. proposed a hybrid Intrusion detection systems based on LLM to solve the intrusion detection problem in the internet of vehicles. Experiments show that the proposed method is able to make up for the shortcomings of traditional methods in several aspects such as classification compared to machine learning based methods. In addition, the proposed method performs well on multiple intrusion detection challenge datasets and shows good generalized recognition ability for in-vehicle networks of different vehicles. In \cite{ref39}, Liu \textit{et al}. modeled a RIS-based IoT communication system using LLM and proposed an optimization strategy for wireless resource allocation, and the simulation results demonstrate the advantages of LLM-enhanced reconfigurable intelligent surface. In \cite{ref40}, Chen \textit{et al}. proposed a dual deep deterministic policy gradient framework under the guidance of LLM to achieve efficient coordination between connected electric vehicles and distributed networks. Experiments show that the proposed method performs well in scheduling tasks compared to traditional methods. All the above literature have conducted a preliminary study on the application of LLM in vehicular environment, but the study on the allocation of vehicular resources and the generation of a queue of tasks for vehicular environment has not been addressed.
	
	As mentioned above, the literature related to digital twin vehicular network has not yet applied LLM whereas the literature that has conducted a preliminary study on LLM has not yet dealt with vehicular resource allocation. Therefore, we try to utilize LLM to conduct research on vehicular resource allocation in digital twin edge network.
	
	\begin{figure}[h]
		\centering
		\includegraphics[width=\columnwidth]{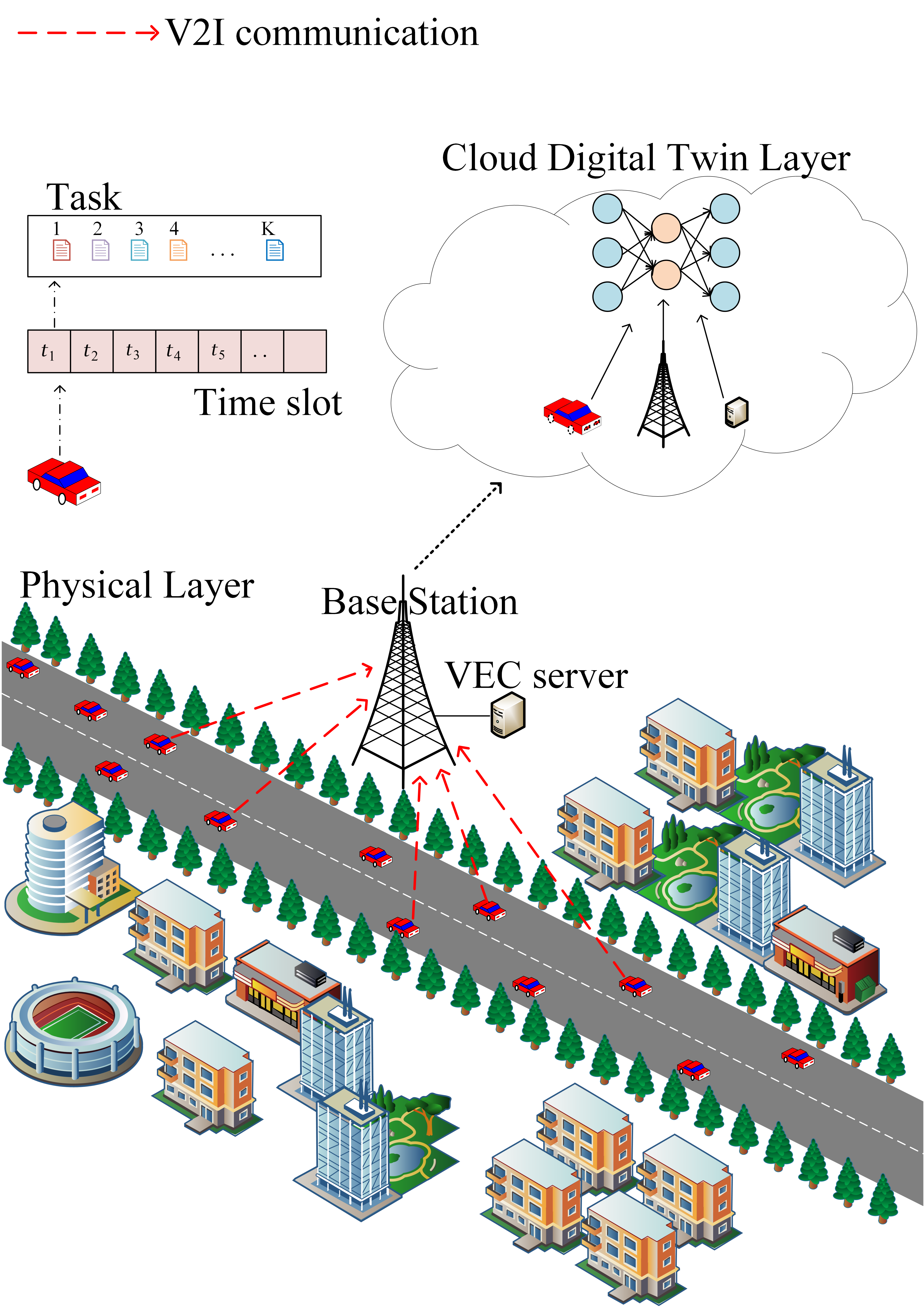}
		\caption{Digital Twin Vehicle Edge Network}
		\label{figure1}
	\end{figure}

	\section{System model}
	Figure 1 shows a system scenario in which we construct a digital twin vehicular networking scenario with $N$ vehicles traveling on the road and a base station equipped with servers. It contains two layers, i.e., the physical layer, which consists of $N$ vehicles, the base station and the server in the scenario, and the cloud digital twin layer, which includes the digital mapping of the entities in the physical layer on the cloud, i.e., the constructed digital copies. Specifically, within the communication coverage of the base station, a vehicle traveling along the road generates $K$ types of tasks to be processed at each time slot. For the sake of this study, we assume that the vehicle always travels within the BS coverage area. Due to limited vehicular resources, the twin model of the vehicle in the cloud twin layer analyzes the specific information about the tasks generated by the vehicle in that time slot, and issues commands to offload each task to the server through transmission with the base station, which has $K$ queues, and accepts the tasks from the vehicle. Then, it stores them in separate dedicated queues before they are processed. In addition, the server is equipped with $N_E$ blocks of CPU cores and has a total resource of $f_E$ cycles per second for all CPUs.
	
	\subsection{Digital Twin Model}
	The digital twin model of vehicle $n$ in the cloud can be represented as
	\begin{equation}
		M^{DT}_{n} = \{I_n, \Gamma _n (t)\},
	\end{equation}
	where $I_{n}$ is configuration information for vehicle $n$, including velocity $v_n$ and position $l_n(t)$,  $\Gamma _n (t)$ is the information about Task of vehicle $n$ in time slot $t$, which denotes $\Gamma_n(t)=\{D_{n,1}^t,D_{n,2}^t,...,D_{n,k}^t,...,D_{n,K}^t\}$. Here, $I_n$ and $\Gamma _n (t)$ are unrelated to each other, as $I_n$ is the current operational state of the vehicle as it travels, while $\Gamma _n (t)$ represents the vehicular tasks at a given time slot.
	
	\subsection{Communication Model}
	Since we deploy the digital twin in the cloud, we ignore the communication time between the twin model and the VEC server. In this scenario, we use orthogonal frequency division multiplexing technique for communication. Therefore, the transmission rate between vehicle $n$ and BS can be expressed as
	\begin{equation}
		r_n ^t = w\log_2 (1+\frac{p_n g_{n,b}^t}{\rho^2}),
	\end{equation}
	where $w$ is the channel bandwidth, $p_n$ is the transmission power of vehicle $n$, $g_{n,b}^t$ is the channel gain between vehicle $n$ and the base station at time slot $t$, and $\rho^2$ is the channel noise between vehicle $n$ and the base station. For the channel gain $g_{n,b}^t$ at the current time slot $t$, it can be modeled as\cite{ref25}
	\begin{equation}
		g_{n,b}^t=|s_{n,b}^t|^2 h_{n,b}^t,
	\end{equation}
	where, $h_{n,b}^t$ is the large-scale fading component, which consists of path fading and shadow fading, where the path fading can be calculated as $128.1+37.6\log_{10}(Dis(V_n,\text{BS}))$, and the shadow fading follows the log normal distribution. $dis(V_n,\text{BS})$ is the distance between vehicle $n$ and BS. $s_{n,b}^t$ is the small-scale channel fading component, which follows the circularly symmetric complex Gaussian with unit variance. To characterize it, we use a first-order Gaussian Markov process for this purpose. Thus, it can be updated in a way that can be expressed as
	\begin{equation}
		s_{n,b}^t=\kappa s_{n,b}^{t-1} + e_{n,b}^t,
	\end{equation}
	where $\kappa$ is the correlation coefficient and $e_{n,b}^t$ is the channel innovation process with distribution $\mathcal{CN}(0,1-\kappa^2)$. The correlation coefficient $\kappa$ can be expressed as $\kappa=J_0 (2\pi f_d \Delta t)$, where $J_0 (\bullet)$ is the zeroth-order Bessel function, and $f_d$ is the maximum Doppler frequency, which can be computed as $f_d = (v_n f_c)/c$, $f_c$ is the carrier frequency, and $c$ is the transmission rate of the electromagnetic wave.
	\subsection{Queue Model}
	At the beginning of time slot $t$, vehicle $n$ generates $K$ types of computing tasks, and thus, we denote the $k$-th type of them as $D_{n,k}^t$. Upon receiving the offloading policy, vehicle $n$ executes the offloading scheme for task $k$. Specifically, the vehicle will offload the $\omega_{n,k}^t |D_{n,k}^t |$ portion of task $k$ to the $k$-th queue at the VEC, $|D_{n,k}^t |$ being the size of task $D_{n,k}^t$. The cumulative amount of the $k$-th type of tasks arriving at queue $k$ within the time interval $\left[t_1,t_2\right)$ can be denoted as $C_k (t_1,t_2 )=\Sigma_{t_1}^{t_2-1}\Sigma_{n=1}^N \omega_{n,k}^t |D_{n,k}^t| $, where $t_1$ and $t_2$ are two adjacent time slots. For the tasks in the queue, the server executes a resource allocation policy to allocate resources to the tasks in each queue. We denote the CPU frequency required to process per bit of data for the $k$-th type of tasks as $c_k$. Thus, the number of $k$-th type tasks processed by the VEC server at time slot $t$ can be expressed as $\phi_k(t)=\frac{f_E \Sigma_{n=1}^N \alpha_{n,k}^t}{c_k}$, where $\alpha_{n,k}^t$ denotes the proportion of resources for the $k$-th task of processing vehicle $n$ and $f_E$ is the total computing resources of the server. Therefore, the backlog of queue $k$ at time slot $t+1$ can be expressed as\cite{ref41},\cite{ref42}
	\begin{equation}
		q_k(t+1)={[q_k(t)+Z_k(t)-\phi_k(t)]}^+,
	\end{equation}
	where $Z_k(t)=\Sigma_{n=1}^N\omega_{n,k}^t|D_{n,k}^t|$ denotes the number of tasks of type $k$ that arrive at queue $k$ at time slot $t$. Since the tasks generated in each time slot enter the queue, maintaining queue stability is essential. If the queue is not kept stable, then the tasks arriving at the queue are at risk of being discarded, which is clearly not in line with the needs of the vehicle itself. Therefore, we denote maintaining strong stability of queue $k$ as\cite{ref43}
	\begin{equation}
		\lim\limits_{t\rightarrow\infty}\text{sup}\frac{1}{t}\sum_{s=0}^{t-1}\sum_{k=1}^{K}\mathbb{E}[q_k(s)]<\infty,
	\end{equation}
	it is important to note that our work focuses on the long-term stability of the queue under this network, which is manifested by ensuring that the length of this queue is stable rather than continuously growing as time increases.

	\subsection{Computing Model}
	\subsubsection{QoS}
	Vehicle $n$ generates tasks and then offloads them according to the offloading policy, for the $k$-th type of tasks, it will be partially offloaded according to $\omega_{n,k}^t$, when $\omega_{n,k}^t=0$, it will be processed directly locally, when $0<\omega_{n,k}^t<1$, it indicates that the task will be partially offloaded for processing, and when $\omega_{n,k}^t=1$, it will be completely offloaded for processing. We let $D_n$ denote the set of tasks generated by vehicle $n$. For the $k$-th type of tasks $D_{n,k}^t$ among them, we denote it as $D_{n,k}^t=\{|D_{n,k}^t |,T_{n,k}^{\text{max}}\}$, where $T_{n,k}^{max}$ is the maximal latency limit of the $k$-th type of tasks generated by vehicle $n$. Therefore, when the $k$-th type of tasks of vehicle $n$ is executed locally, its latency can be calculated as
	\begin{equation}
		t_{n,k}^{\text{local}}=\frac{(1-\omega_{n,k}^t)|D_{n,k}^t|c_k}{f_v},
	\end{equation}
	where $f_v$ is the vehicle's own computing resources. For the offloading part of the task, it first needs to be transmitted to the server, and after it enters the server's queue, the server executes a resource allocation policy to allocate CPU frequencies to process it. In addition, it should be noted that due to the mapping difference between the digital twin model in the cloud and its real physical entity, the CPU frequency allocated by the VEC server to the $k$-th class task of vehicle $n$ actually has an estimated bias $\Delta f_{n,k}^{est}$ \cite{ref26}, which in turn has a bias in the latency of the final processing of the task, which is denoted as $\Delta t_{n,k}^{est}$, which can be calculated as
	\begin{equation}
		\Delta t_{n,k}^{\text{est}}=-\frac{\omega_{n,k}^t|D_{n,k}^t|c_k\Delta f_{n,k}^{\text{est}}}{(\alpha_{n,k}^t f_E)(\Delta f_{n,k}^{\text{est}}+\alpha_{n,k}^t f_E)},
	\end{equation}
	therefore, for the delay $t_{n,k}^{\text{edge}}$ obtained after edge processing, it can be finally represented as
	\begin{equation}
		t_{n,k}^{\text{edge}}=\frac{|D_{n,k}^t|}{r_n^t}+\frac{\omega_{n,k}^t|D_{n,k}^t|c_k}{\alpha_{n,k}^t f_E}+\Delta t_{n,k}^{\text{est}}.
	\end{equation}
	For completing the task $D_{n,k}^t$, it needs to complete both local and offload processing, so its final processing delay can be expressed as
	\begin{equation}
		t_{n,k}=t_{n,k}^{\text{local}}+t_{n,k}^{\text{edge}}.
	\end{equation}
	To measure the relationship between the delay for completing a class $k$ task for vehicle $n$ and its maximum delay limit, we express it in terms of QoS, which can be calculated as\cite{ref44}
	\begin{equation}
		U_{n,k}(t)=1-\frac{t_{n,k}}{T_{n,k}^{\text{max}}},
	\end{equation} 
	since the number of tasks generated by vehicle $n$ in a single time slot is $K$, the average QoS for vehicle $n$ is expressed as
	\begin{equation}
		U_n^{\text{ave}}(t)=\frac{1}{K}\sum_{k=1}^{K} U_{n,k}(t),
	\end{equation}
	thus, for this network, the average QoS of the whole system can be expressed as
	\begin{equation}
		U^{\text{sys}}(t)=\frac{1}{N}\sum_{n=1}^{N}U_n^{\text{ave}}(t),
	\end{equation}
	
	\subsubsection{Energy}
	In addition to focusing on the QoS affected by processing delay under this system, the energy consumption of the system is equally important. Under this system, the total energy consumption is mainly divided into local processing energy and edge processing energy. For the local processing energy consumption, it is mainly generated when the vehicle's own computing resources are used to process the local computing tasks, which is calculated as follows
	\begin{equation}
		E^{\text{local}}(t)=\sum_{n=1}^{N}\sum_{k=1}^{K}{f_v}^3	\kappa^{\text{ve}} t_{n,k}^{\text{local}},
	\end{equation}
	where  $\kappa^{\text{ve}}$ denotes the effective switching capacitance of the computing device on the vehicle. For the calculation of the edge energy consumption, since it mainly involves the consumption of the CPU power of the server, based on the dynamic frequency regulation method that is widely used to construct the real CPU power consumption, the edge energy consumption is calculated as\cite{ref41}
	\begin{equation}
		E^{\text{edge}}(t)=N_E\kappa^{\text{se}}(\frac{f_E\Sigma_{n=1}^N\Sigma_{k=1}^K\alpha_{n,k}^t}{N_E})^3,
	\end{equation}
	where $\kappa^{\text{se}}$ denotes the effective switching capacitance parameter of the server hardware, which has the server hardware itself to decide. For the total energy consumption of the system under this network, it consists of the local energy consumption and the edge energy consumption, so we can get
	\begin{equation}
		E^{\text{sys}}(t)=E^{\text{local}}(t)+E^{\text{edge}}(t),
	\end{equation}
	\subsection{Optimization Problem}
	Our goal is to maximize the average QoS of the system under this network while minimizing the total energy consumption of the system as much as possible, therefore, we construct the following optimization problem:
	\begin{subequations}\label{P1}
		\begin{equation}
			P1:\mathop {\min}_{(\omega_{n,k}^t,\alpha_{n,k}^t)} -U^{\text{sys}}(t) + E^{\text{sys}}(t)\quad\quad
		\end{equation}
		\begin{equation}
			\qquad s.t.\lim\limits_{t\rightarrow\infty}\text{sup}\frac{1}{t}\sum_{s=0}^{t-1}\sum_{k=1}^{K}\mathbb{E}[q_k(s)]<\infty
		\end{equation}
		\begin{equation}
			t_{n,k}\leq T_{n,k}^{\text{max}}
		\end{equation}
		\begin{equation}
			\qquad\Sigma_{n=1}^N\Sigma_{k=1}^K \alpha_{n,k}^t f_E + \Delta f_{n,k}^{\text{est}}\leq f_E
		\end{equation}
		\begin{equation}
			\omega_{n,k}^t \in [0,1]
		\end{equation}
	\end{subequations}
	
	In particular, constraint (17b) indicates they need to ensure the stability of the queue in order to avoid tasks being discarded, constraint (17c) indicates that the latency obtained from completing the task must not exceed the maximum latency limit, constraint (17d) indicates that the computational resources consumed by the offload processing portion of all vehicle-generated tasks must not exceed the maximum amount of resources of the VEC server itself, and constraint (17e) indicates a range of values for the offloading policy. Note that constraint (17b) is affected by the offloading decision, which is interconnected with the resource allocation decision, and both affect constraint (17c). Thus, the complexity of this interaction makes it difficult to solve problem $P1$ using traditional operations. In order to overcome this suffering, we transformed the problem constraints and reconstructed the problem using Lyapunov optimization.
	
	\subsection{Lyapunov Optimization}
	In this section, we will use Lyapunov optimization to pair transform the long-run stability problem of the queue into a short-run problem. Let $\Delta Q(t)$ denote the conditional Lyapunov shift at time slot $t$, which is represented as
	\begin{equation}
		\Delta Q(t)\stackrel{\triangle}{=}E\{L(Q(t+1))-L(Q(t))|Q(t)\},
	\end{equation}
	where $Q(t)=[q_1(t), q_2(t), ..., q_K(t)]$ is the backlog of the queue at time slot $t$. Let $L(Q(t))$ denote the Lyapunov function used to measure the average queue backlog, which is denoted as $L(Q(t))=\frac{1}{2}\Sigma_{k=1}^K(q_k(t))^2$. Squaring both sides of Eq. (5) and bringing the Lyapunov function Eq. (18), then
	\begin{equation}
		\begin{split}
			\Delta Q(t) \le&\frac{1}{2}E\{\Sigma_{k=1}^K(Z_k(t)-\phi_k(t))^2|Q(t)\}+\\
			&E\{\Sigma_{k=1}^Kq_k(t)(Z_k(t)-\phi_k(t))|Q(t)\},
		\end{split}
	\end{equation}
	Let $Z_k^{\text{max}}$ serves as an upper bound for $Z_k(t)$, then $Z_k(t)\le Z_k^{\text{max}}$ can be obtained. Moreover, since $\Sigma_{n=1}^N \alpha_{n,k}^t\le1$ in $\phi_k(t)=\frac{f_E\Sigma_{n=1}^N \alpha_{n,k}^t}{c_k}$, according to the drift plus penalty bound\cite{ref45}, the first term on the right-hand side of Eq. (19) can be upper bounded by
	\begin{equation}
		\begin{aligned}
			\frac{1}{2}&E\{\Sigma_{k=1}^K(Z_k(t)-\phi_k(t))^2|Q(t)\}\le\\ &\frac{1}{2}E\{\Sigma_{k=1}^K(Z_k(t)^2-\phi_k(t)^2)|Q(t)\}\\
			&=\frac{1}{2}E\{\Sigma_{k=1}^K (Z_k^{max})^2-(\frac{f_E}{c_k})^2|Q(t)\}\\
			&=B
		\end{aligned}
		\label{1}.
	\end{equation}
	Bringing the result into Eq. (19), then
	\begin{equation}
		\begin{aligned}
			\Delta &Q(t)\le\\ &B+E\{\Sigma_{k=1}^Kq_k(t)(Z_k(t)-\frac{f_E\Sigma_{n=1}^N\alpha_{n,k}^t}{c_k})|Q(t)\}
		\end{aligned}
		\label{2}.
	\end{equation}
	
	We then employ the opportunistic minimization of expectations technique for the second term on the right-hand side of Eq. (21), after which we obtain the following result
	\begin{equation}
		\Delta Q(t)\le B-\Sigma_{k=1}^K q_k(t)(\frac{f_E\Sigma_{n=1}^N \alpha_{n,k}^t}{c_k}-Z_k(t)).
	\end{equation}
	
	Let $\frac{f_E\Sigma_{n=1}^N \alpha_{n,k}^t}{c_k}-Z_k(t)$ be denoted by $\epsilon$, so we finally rewrite Eq. (22) as $\Delta Q(t)\le B-\epsilon\Sigma_{k=1}^K q_k(t)$. According to conditional Lyapunov shift theory, if the upper bound on the Lyapunov shift $\Delta Q(t)$ is $B-\epsilon\Sigma_{k=1}^K q_k(t)$ for each time slot $t$, we can rewrite the constraints(16b) in Problem 1 as
	\begin{equation}
		\lim\limits_{t\rightarrow\infty}sup\frac{1}{t}\sum_{s=0}^{t-1}\sum_{k=1}^{K}\mathbb{E}[q_k(s)]\le\frac{B}{\epsilon}.
	\end{equation}
	
	In this scenario, to maintain the stability of the queue, it is sufficient to ensure that the conditional Lyapunov shift has a minimal tight upper bound, i.e., there exists an optimal $a_{n,k}(t)$ to minimize the right-hand side of Eq. (22). Combining this short-term decision problem with problem $P1$, we get problem $P2$:
	\begin{subequations}\label{P2}
		\begin{equation}
			\begin{aligned}
				P2:\mathop {\min}_{(\omega_{n,k}^t,\alpha_{n,k}^t)} \beta\Sigma_{k=1}^K q_k(t)(Z_k(t)-\frac{f_E\Sigma_{n=1}^N\alpha_{n,k}^t}{c_k}) \\
				+(-U^{\text{sys}}(t) + E^{\text{sys}}(t))\quad\quad
			\end{aligned}
		\end{equation}
		%\begin{equation}
		%	\qquad s.t.\lim\limits_{t\rightarrow\infty}sup\frac{1}{t}\sum_{s=0}^{t-1}\sum_{k=1}^{K}\mathbb{E}[q_k(s)]\le\frac{B}{\epsilon}
		%\end{equation}
		\begin{equation}
			t_{n,k}\leq T_{n,k}^{\text{max}}
		\end{equation}
		\begin{equation}
			\qquad\Sigma_{n=1}^N\Sigma_{k=1}^K \alpha_{n,k}^t f_E + \Delta f_{n,k}^{\text{est}}\leq f_E
		\end{equation}
		\begin{equation}
			\omega_{n,k}^t \in [0,1]
		\end{equation}
	\end{subequations}
	where $\beta$ is a non-negative tradeoff factor.
	
	The reconstructed problem $P2$ solves the long-run decision-making problem, but the problem is still a complex optimization problem. We choose to utilize LLM-based in-context learning to solve the problem, but before we can do so, we need to obtain a example set which got after training through the MARL method.
	
	\begin{figure*}[h]
		\centering
		\includegraphics[width=\textwidth]{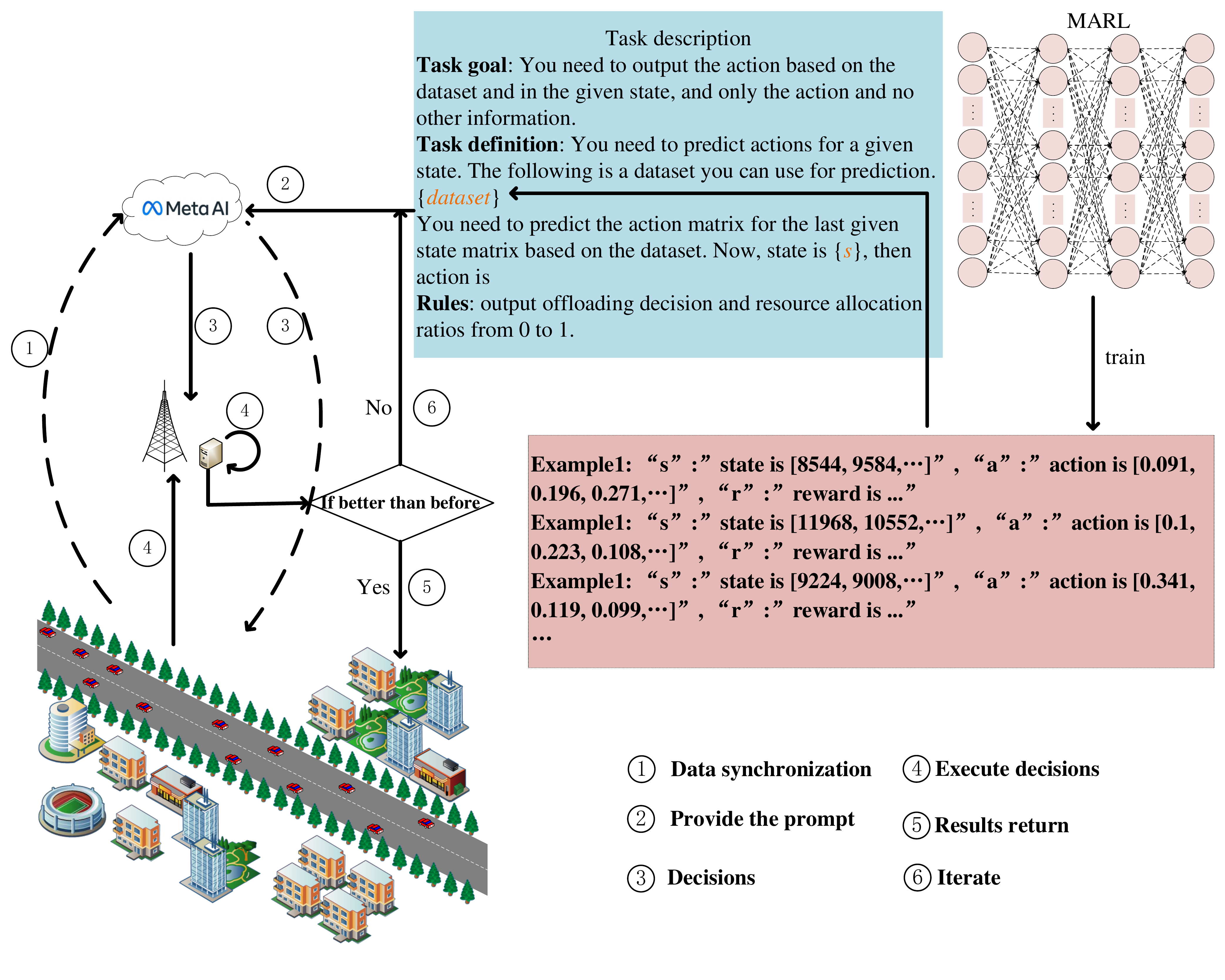}
		\caption{Overall design of LLM-based context learning}
		\label{figure2}
	\end{figure*}
	
	\section{Proposed Solution}
	To address the optimization problem, we first employ MARL to obtain an initial set of cases, and then utilize LLM to optimize the actions learned from this case set.
	\subsection{MARL-Based Case Set}
	The framework required for MARL, i.e. state, action, reward function, needs to be constructed.
	\subsubsection{State}
	In this setting, we let the state consist of the task $\Gamma_n(t)$, vehicle position $l_n(t)$, vehicle speed $v_n$, channel gain $g_{n,b}^t$, i.e.
	\begin{equation}
		s_n(t) = \{\Gamma_n(t), l_n(t), v_n, g_{n,b}^t\},
	\end{equation}
	therefore, the environment state is $S(t)=\{s_n(t)\}_N$.
	\subsubsection{Action}
	We take the offloading ratio $\omega_{n,k}^t$ and the resource ratio $a_{n,k}^t$ as the actions executed by the agent $n$ after obtaining the observation, i.e.
	\begin{equation}
		a_n(t) = \{\omega_{n,k}^t, \alpha_{n,k}^t\},
	\end{equation}
	after that, we can get the overall action, which is represented as $A(t)=\{a_n(t)\}_N$.
	\subsubsection{Reward}
	After considering various constraints, the reward function is expressed as
	\begin{equation}
		r_n(t)=U_n^{ave}-\eta(f_E - (\Sigma_{n=1}^N\Sigma_{k=1}^K a_{n,k}^t f_E + \Delta f_{n,k}^{\text{est}})), 
	\end{equation}
	where $\eta$ is a weight parameter used to keep the two terms on the right-hand side of the equation to an order of magnitude. Then, the long-term discount reward is $R_n(t)=\Sigma_{t_0}^t \gamma_n r_n(t)$, where $t_0$ is the previous time, $\gamma_n$ is the discount factor. We can achieve the optimal action by maximizing the long-term discount reward of each agent.
	
	The algorithm procedure follows the implementation in \cite{ref26}. Among the neural networks involved in the algorithm are the actor network, critic network and their respective target networks. Their specific structure is as follows:
		\subsubsection*{Actor}
		We categorize the actor network into three layers: the input layer, the fully connected layer, and the output layer. The fully connected	layer consists of three hidden layers and a softmax layer. The first two hidden layers use the rectified linear unit (ReLu) function as an activation function and the third hidden layer uses the tangent (tanh) function as an activation function. Since the actor network is divided into two parts, we will explain the estimated actor network and the target actor network separately.
		
		The input is the current state for the estimated actor network, including the computing tasks generated by the vehicle, the vehicle position, the vehicle speed and the channel gain. The current state is passed through three hidden layers after outputting the probability of possible actions. Then after passing through the softmax layer, the total probability of all actions is set to 1. Finally, agent $n$ randomly selects one of the outputs.
		
		\subsubsection*{Critic}
		The structure of the critic network also includes an input layer, a fully connected layer and	an output layer. The difference is that there is no softmax layer in the fully connected layer, only three hidden layers, and their activation functions are Relu, Relu and Tanh respectively.
		
		The inputs to the estimation critique network are the states and actions of all agents at the current epoch, and the output is the Q value.
	
	%where both the actor and critic networks adopt a three-hidden-layer architecture.  Specifically, the actor network employs rectified linear unit (ReLu), ReLu, and hyperbolic tangent (Tanh) activation functions for its hidden layers respectively, while the critic network uses ReLu activation functions throughout all hidden layers.
	
	After the case set obtained through MARL training we use the LLM based on in-context learning to learn the case set and then complete the decision-making for the randomly arriving real-world situations.
	
	\subsection{In-context Learning-based Approach}
	In-context learning means that LLM can learn from human characters, such as task descriptions and task solution cases, in order to improve the performance of the target task. Context learning can be defined as\cite{ref35}
	\begin{equation}
		D_{task}\times\epsilon_{example}^t\times s_t\times\mathcal{LLM}\Rightarrow A_t,
	\end{equation}
	where $D_{task}$ is the task description, $\epsilon^t_{example}$ is a collection of some cases under time slot $t$, $s_t$ is the environment state under the current time slot $t$, which is related to the target task, $\mathcal{LLM}$ is the LLM model, as well as $A_t$ is the output result of LLM. Specifically, by inputting a description $D_{task}$ of the task we need to solve to the LLM, the LLM learns from the set of cases $\epsilon_{example}^t$, and then makes a decision $A_t$ on the target task under the current time slot $t$ based on the current state of the environment $s_t$. In the following, we will make a decision $A_t$ on the task description $D_{task}$ and the set of cases $\epsilon_{example}^t$ are described.
	
	\subsubsection{Task Description}
	$D_{task}$ is important for LLM to provide key information about the problem to be solved. Typically, a task description $D_{task}$ consists of three parts, which are task goal, task definition, and additional rules. Below is the prompt we designed to inspire the LLM, which contains a specific description of the task.
	
	\begin{userbox}
		You are a mathematical tool to predict some models. You need to predict actions for a given state. The following is a dataset you can use for prediction. You need to predict the action matrix for the last given state matrix based on the dataset. Please output the action matrix directly without any other information. \texttt{(data set)}
	\end{userbox}
	
	\begin{aibox}
		[[0.091, 0.196, 0.271, 0.038, 0.038, 0.038], 
		[0.1, 0.223, 0.108, 0.038, 0.038, 0.038], 
		[0.203, 0.101, 0.109, 0.038, 0.0325, 0.038], 
		[0.208, 0.113, 0.144, 0.0225, 0.038, 0.038], 
		[0.01, 0.01, 0.99, 0.011, 0.005, 0.005], 
		[0.01, 0.99, 0.01, 0.0065, 0.005, 0.005], 
		[0.162, 0.08, 0.074, 0.038, 0.038, 0.038], 
		[0.105, 0.117, 0.247, 0.038, 0.038, 0.0365],
		[0.341, 0.119, 0.099, 0.038, 0.038, 0.038], 
		[0.01, 0.01, 0.01, 0.0085, 0.038, 0.005]]
	\end{aibox}

	For the task goal, it specifically refers to \textbf{“you need to output the action based on the dataset and in the given state, and only the action and no other information”}. Then we set the task definition to output dimension-specific actions in a given environment state, e.g., when the number of vehicles in the environment changes, the LLM must take this situation into account. After that the case set is represented as \textbf{“Here is a dataset you can use for prediction”}. Finally, we set up additional response rules. That is "an offloading decision and resource allocation task", where the goal is to output offloading decision and resource allocation ratios from 0 to 1, because the former requires LLM to focus on the decision-making process, while the latter facilitates subsequent data extraction.
	
	The above task description provides a template for the definition of optimization tasks through formatted natural language, which avoids the complexity that would arise from dedicating a specific model for this purpose. Moreover, it is also user-friendly, as the operator only needs to make simple additions or deletions to the task descriptions without any specialized knowledge of optimization.
	
	Overall, the design of prompt requires the abstraction of the optimization problem into natural language. The body of the prompt to address the context of the optimization problem, then the purpose of the cue needs to be designed, in addition to adding some guiding words in order to allow LLM to understand the problem to be solved in the context of the case set.
	
	\subsubsection{Case Set}
	Examples are also critical for LLM in contextual learning, so the selection of cases must be careful. This is because on the one hand cases are an important factor in LLM decision making, and LLM relies on cases to justify its decisions, and on the other hand because the invocation of LLM models usually has a token limitation, which means that it is not possible for us to send a large number of examples. Moreover, since the state of the environment in the scenario of this paper is continuous, which means that the number of cases is infinite, picking the most relevant and functional cases is challenging. We choose the set of examples converged by MARL training in the previous section. It should be noted that the case set obtained by MARL is only preliminary, and it needs to be further optimized by LLM based on the cases in it.

	\subsubsection{LLM-based context learning}
	Figure 2 shows the general design of LLM-based context learning. Specifically, when a vehicle in the physical layer generates a class $K$ task at time slot $t$, it means that the information of the vehicle has changed, and therefore, it will synchronize the information to the twin model in the cloud digital twin layer via intra-twin communication. Then, the twin model will invoke the LLM, i.e., it will input the prompt into the LLM and the LLM will make a decision for the new task state based on the historical dataset and it will make two types of decisions, i.e., task offloading decision and resource allocation decision. After making the decisions, the cloud will return these two types of decisions to the vehicle and the server respectively. After receiving the offloading decision, the vehicle will offload the generated task to the server's queue. The server, on the other hand, will calculate and process the resource allocation for the tasks in the queue according to the resource allocation decision, and the result of the processing will be returned to the vehicle. At the same time, it will be added to the dataset as a new example.
	
	\section{Simulation results}
	
	\begin{table}[h]
		\caption{Environmental parameters\label{tab:table1}}
		\renewcommand\arraystretch{1.2}
		\centering
		\resizebox{\columnwidth}{!}{
		\begin{tabular}{|>{\centering\arraybackslash}p{0.2\columnwidth}|>{\centering\arraybackslash}p{0.3\columnwidth}|>{\centering\arraybackslash}p{0.2\columnwidth}|>{\centering\arraybackslash}p{0.3\columnwidth}|}
			\hline
			\textbf{Parameter} & \textbf{Value} & \textbf{Parameter} & \textbf{Value} \\
			\hline
			$w$ & 20MHZ & $p_n$ & 200mW \\
			\hline
			$\rho^2$ & -110mdB & $f_c$ & 2GHZ \\
			\hline
			$f_v$ & 5GHZ & $v_n$ & [10,15]m/s \\
			\hline
			$c$ & $3\times10^8$m/s & $f_E$ & 400GHZ\\
			\hline
			$|D_{n,k}^t |$ & [1000, 1500]Byte & $c_k$ & 0.25MHZ/Byte\\
			\hline
			$\kappa^{ve}$ & $10^{-28}$ & $\kappa^{se}$ & $\frac{1}{(400GHZ)^3}$ \\ 
			\hline
			$N_E$ & 10 & $\beta$ & 1 \\
			\hline
			$\gamma_n$ & 0.95 & $\lambda$ & 0.01\\
			\hline
		\end{tabular}}
	\end{table}

	\subsection{Experimental Setup}
	The simulation environment in this paper is implemented under Python 3.11.0 and the LLM model used is Llama 3.1-8b. In terms of the environment setup, the parameters are set as shown in Table 1. Specifically, the bandwidth $w$ is set to 20 MHZ, the transmission power $p_n$ is 200 mW, the channel noise $\rho^2$ is -110 mdB, and the carrier frequency$f_c$ is 2 GHZ. In addition, the amount of the vehicle's own computing resources $f_v$ is 5 GHZ, the velocity of the vehicle $v_n$ is randomly chose in [10, 15]m/s, the transmission rate of the electromagnetic is $3\times10^8$m/s and the total amount of resources of the edge servers$f_E$ is 400 GHZ. The task size $|D_{n,k}^t|$ is randomly chose in [1000,1500]Byte and the frequency required per bit of the data $c_k$ is 0.25MHZ/Byte. The effective switching capacity of the vehicle's vehicular devices $\kappa^{ve}$ is $10^{-28}$, and that of the servers hardware has an effective switching capacity $\kappa^{se}$ of $\frac{1}{(400\text{GHZ})^3}$, and the number of CPUs on the server $N_E$ is 10. In the experimental part, we set the non-negative coefficient $\beta$ in Problem 2 to 1. The discount factor $\gamma_n$ is 0.95 and the updated rate $\lambda$ is 0.01.
	
	\begin{figure}[!h]
		\centering
		\begin{subfigure}[overall comparison]{
				\centering
				\includegraphics[width=\columnwidth]{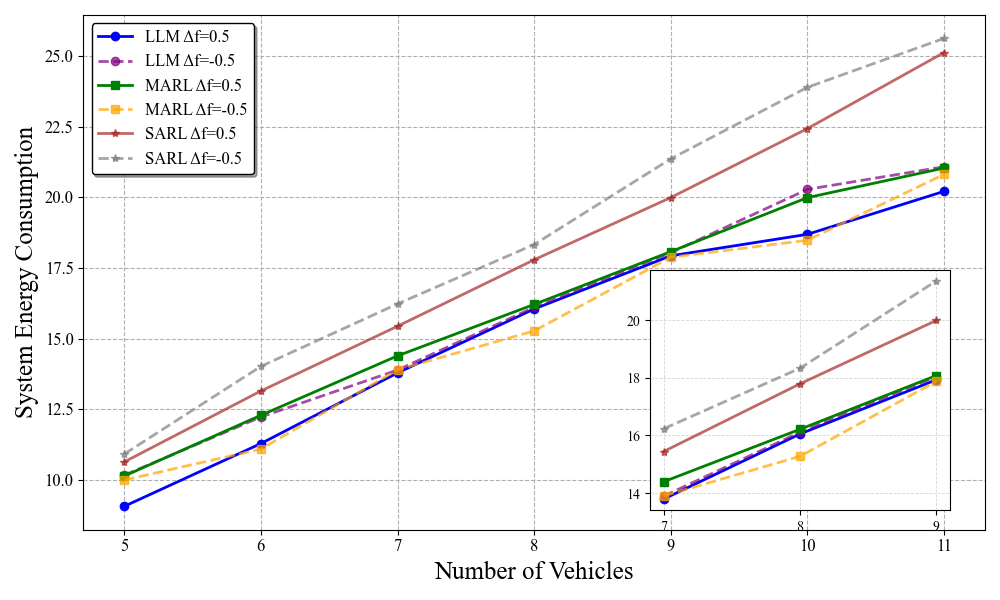}
				%\caption{dt}
				\label{figure3.a}}
		\end{subfigure}
		\hfill
		\begin{subfigure}[Comparison between LLM and MARL]
			{
				\centering
				\includegraphics[width=\columnwidth]{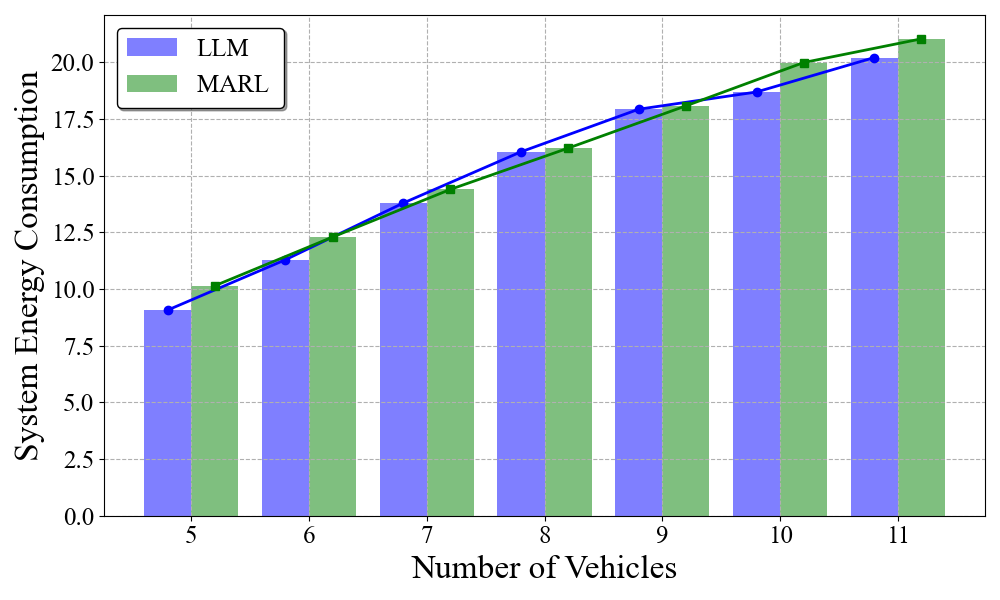}
				%\caption{tk}
				\label{figure3.b}}
		\end{subfigure}
		\hfill
		\begin{subfigure}[Comparison of LLM under different estimation biases]
			{
				\centering
				\includegraphics[width=\columnwidth]{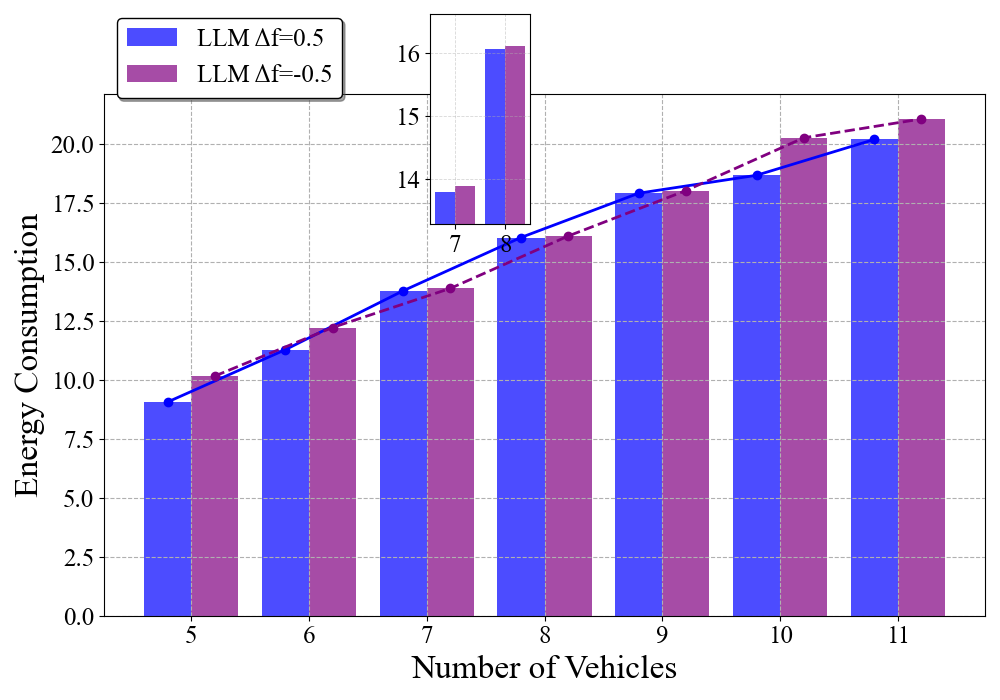}
				%\caption{tk}
				\label{figure3.c}}
		\end{subfigure}
		\caption{Comparison of system energy consumption between LLM, SARL and MARL with different estimation bias.}
		\label{figure3}
	\end{figure}
	
	%fig4
	\begin{figure}[!h]
		\centering
		\includegraphics[width=\columnwidth]{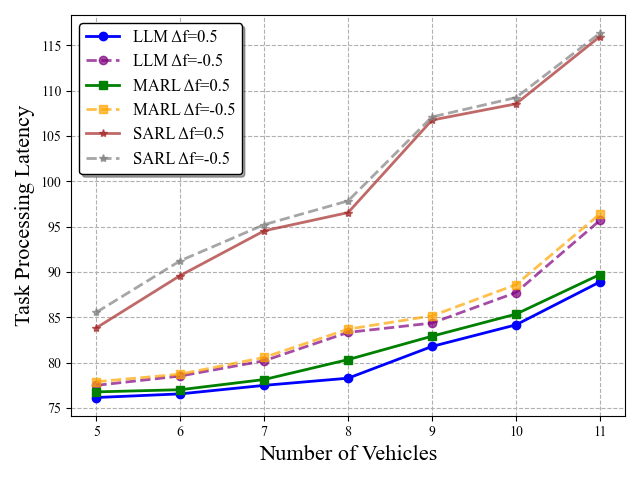}
		\caption{Comparison of task processing latency for LLM, SARL, and MARL with different estimation biases.}
		\label{figure4}
	\end{figure}
	
	%fig5
	\begin{figure}[!h]
		\centering
		\includegraphics[width=\columnwidth]{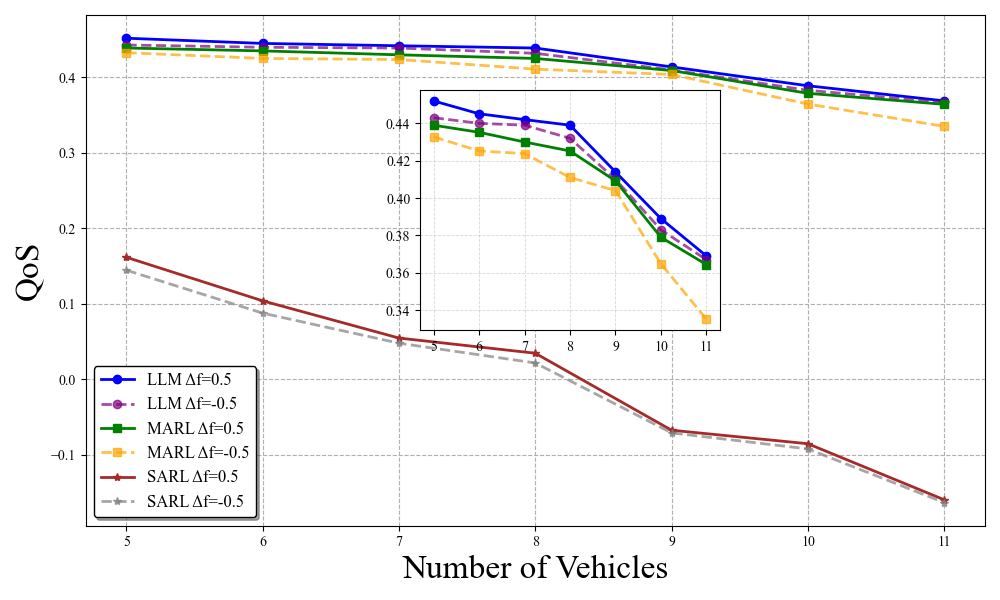}
		\caption{Comparison of QoS of LLM, SARL and MARL with different estimation biases.}
		\label{figure5}
	\end{figure}
	
	%fig6
	\begin{figure}[!h]
		\centering
		\begin{subfigure}[$\Delta f= 0.5$]{
				\centering
				\includegraphics[width=\columnwidth]{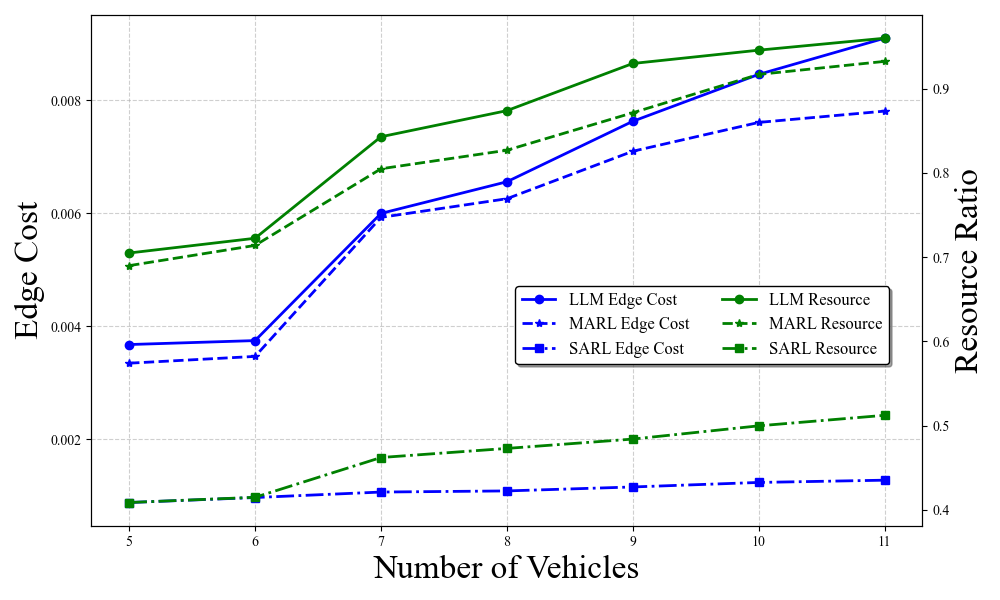}
				%\caption{dt}
				\label{figure6.a}}
		\end{subfigure}
		\centering
		\begin{subfigure}[$\Delta f=-0.5$]
			{
				\centering
				\includegraphics[width=\columnwidth]{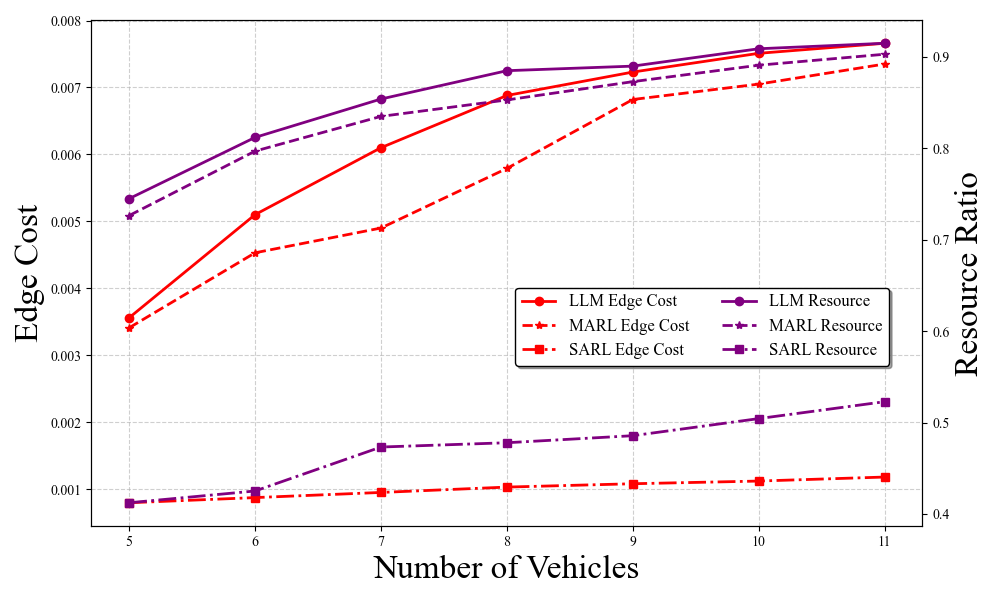}
				%\caption{tk}
				\label{figure6.b}}
		\end{subfigure}
		\caption{Comparison of Edge Energy and Resource Ratios for LLM, SARL, and MARL with Different Estimation Biases.}
		\label{figure6}
	\end{figure}
	
	\subsection{Baseline Algorithm}
	%We consider MARL to be the optimal baseline due to the fact that MARL techniques have been widely used in optimization problems for various types of networks. In addition, considering the scenario of this paper, we choose the MARL as a comparison. 
	We compared with current popular deep reinforcement learning algorithms, i.e. multi-agent or single-agent reinforcement learning (SARL). For MARL, the number of hidden layers as well as the activation function are the same as described in Section IV.B. For SARL, the number of hidden layers is the same but the activation function of the neural network is ReLu function.
	
	\subsection{Performance Evaluation}
	Fig. 3 shows the system energy consumption of LLM, MARL, and SARL for different twin mappings. In Fig. 3(a), it can be seen that for all the discussed methods, the system energy consumption increases with the number of vehicles, where SARL has the highest energy consumption while LLM is able to achieve similar results as MARL. The energy consumption of SARL is higher since making decisions based on local information, having relatively weak capability for task offloading and resource scheduling. In order to be able to make a more specific comparison, we can get through Fig.3.b that the energy consumption achieved by MARL is higher than that obtained through LLM. On the other hand, in order to compare the situation under different estimation biases, we see through Fig.3.c that the energy consumption achieved by LLM is lower when the bias is positive than when it is negative. This is because, when the bias is positive, the amount of resources estimated by the digital twin will be higher, and therefore, less resources will be subsequently invested, but on the contrary, more resources will be invested.
	%Figure 3 exhibits the system energy consumption achieved by LLM and MARL for different twin estimation biases, respectively. It can be seen that the system energy consumption increases with the increase in the number of vehicles for both methods. In Fig.3.a, we can see that LLM is able to achieve a performance comparable to the popular MARL algorithm. 

	Figure 4 compares the task processing latency of LLM, SARL, and MARL for different estimation biases. It can be seen that the task processing delay increases with the number of vehicles. In addition, SARL, as mentioned in the previous section, makes decisions based on local information and thus fails to obtain the optimal scheduling plan, i.e., SARL cannot adapt to the environment of intense competition for resources. In contrast, LLM achieves better results than MARL because LLM performs comprehensive learning of the environment based on a typical set of cases and makes smart inference decisions on new cases. On the other hand, the delay when the bias is positive is lower than when it is negative. This is due to the fact that, as mentioned earlier, the amount of resources estimated by the digital twin is higher when the bias is positive, which naturally leads to lower latency.
	%Figure 4 compares the task processing latency of LLM and MARL under different estimation biases. It can be seen that the task processing delay increases with the number of vehicles. In addition, LLM achieves better performance than MARL.This is 

	Fig. 5 shows the comparison of LLM, SARL and MARL in terms of QoS. We can get that QoS decreases as the number of vehicles increases. In this case, the negative value of QoS for SARL is due to the fact that SARL's task processing delay is higher than the maximum delay tolerance due to poor scheduling scheme in case of higher number of vehicles as shown in Fig. 4. On the other hand, even though the QoS is also decreasing, the performance achieved by LLM is better than MARL.
	%Figure 5 shows the comparison between LLM and MARL in terms of QoS. We can get that QoS decreases as the number of vehicles increases. This is because as shown in Fig. 4, the delay is pushed up with the increase in the number of vehicles, which naturally leads to a decrease in the QoS obtained from Eq.11. 
	
	Fig. 6 compares the edge energy and resource ratios of LLM, SARL, and MARL for different estimation biases. It can be seen that SARL achieves the lowest edge energy consumption, which is due to its relatively weak ability to schedule resources as well as offload tasks, as mentioned earlier. In addition, LLM is higher than MARL in terms of edge consumption and resource share, regardless of whether the deviation is positive or negative. This is because, as illustrated in Fig. 4, LLM occupies more computing resources in order to get lower processing latency, which leads to an increase in the resource share, and occupying more computational resources leads to higher edge energy consumption.
	
	\section{Conclusions}
	In this paper, we considered the problem of generating multiple tasks by vehicles in a single time slot and queuing of tasks in the server under digital twin edge networks, while constructing a preliminary optimization problem by analyzing this situation. Since the preliminary constructed problem has a long term decision constraint that is difficult to solve, we used Lyapunov optimization to transform this condition into a short term decision. After constructing the new problem, we utilized the LLM method, which is not widely used in the field of IoT today, to solve the problem. Experimental results have shown that the LLM-based In-context learning method can achieve similar or even better performance than the widely popular MARL. However, the method proposed in this paper also has limitations, such as the speed of operation of LLM is dependent on the performance of the hardware deployed. Therefore, there is a requirement for the deployment cost. In addition, overly complex optimization problem is a challenge for LLM. Therefore, in the future work, we will try to decompose the complex problem into several sub-problems and hand them over to LLM for processing. At the same time, we will also consider lightweight models to minimize deployment costs. We summarize them as follows:
	\begin{itemize}
		\item The LLM can understand the tasks to be accomplished and the objectives of the tasks to be achieved through the cue words we designed.
		\item The LLM is able to learn the characteristics of the task to be accomplished from a typical case set and can make reasoned decisions for new tasks.
	\end{itemize}

	\begin{IEEEbiography}[{\includegraphics[width=1.1in,height=1.4in,clip,keepaspectratio]{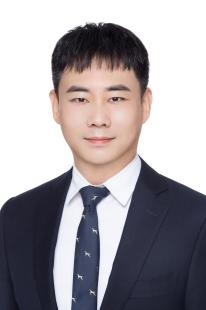}}] {Qiong Wu} (Senior Member, IEEE) received the Ph.D. degree in information and communication engineering from National Mobile Communications Research Laboratory, Southeast University, Nanjing, China, in 2016. From 2018 to 2020, he was a postdoctoral researcher with the Department of Electronic Engineering, Tsinghua University, Beijing, China. He is currently an associate professor with the School of Internet of Things Engineering, Jiangnan University, Wuxi, China. 
		
	Dr. Wu is a Senior Member of IEEE and China Institute of Communications. He has published over 80 papers in high impact journals and conferences, and authorized over 30 patents. He was elected as one of the world's top 2\% scientists in 2024 and 2022 by Stanford University. He has received the young scientist award for ICCCS'24 and ICITE’24. He won the high-impact paper of Chinese Journal of Electronics award. He has been awarded the National Academy of Artifical Intelligence (NAAI) Certified AI Senior Engineer, and was the excellent reviewer for Computer Networks in 2024. He has severed as the editorial board member of Sensors and CMC-Computers Materials \& Continua, the early career editorial board member of Radio Engineering and Chinese Journal on Internet of Things, the lead guest editor of Sensors, CMC-Computers Materials \& Continua, Radio Engineering and Frontiers in Space Technologies, the guest editor of Electronics and Chinese Journal on Internet of Things, the TPC co-chair of WCSP'22, the workshop chair of NCIC'23/25, ICFEICT'24/25, CIoTSC’24, IAIC’24, RFAT‘25 and FRSE’25, as well as the TPC member and session chair for over 10 international Conferences. His current research interest focuses on vehicular networks, autonomous driving communication technology, and machine learning.
	\end{IEEEbiography}
	
	\begin{IEEEbiography}[{\includegraphics[width=1.1in,height=1.4in,clip,keepaspectratio]{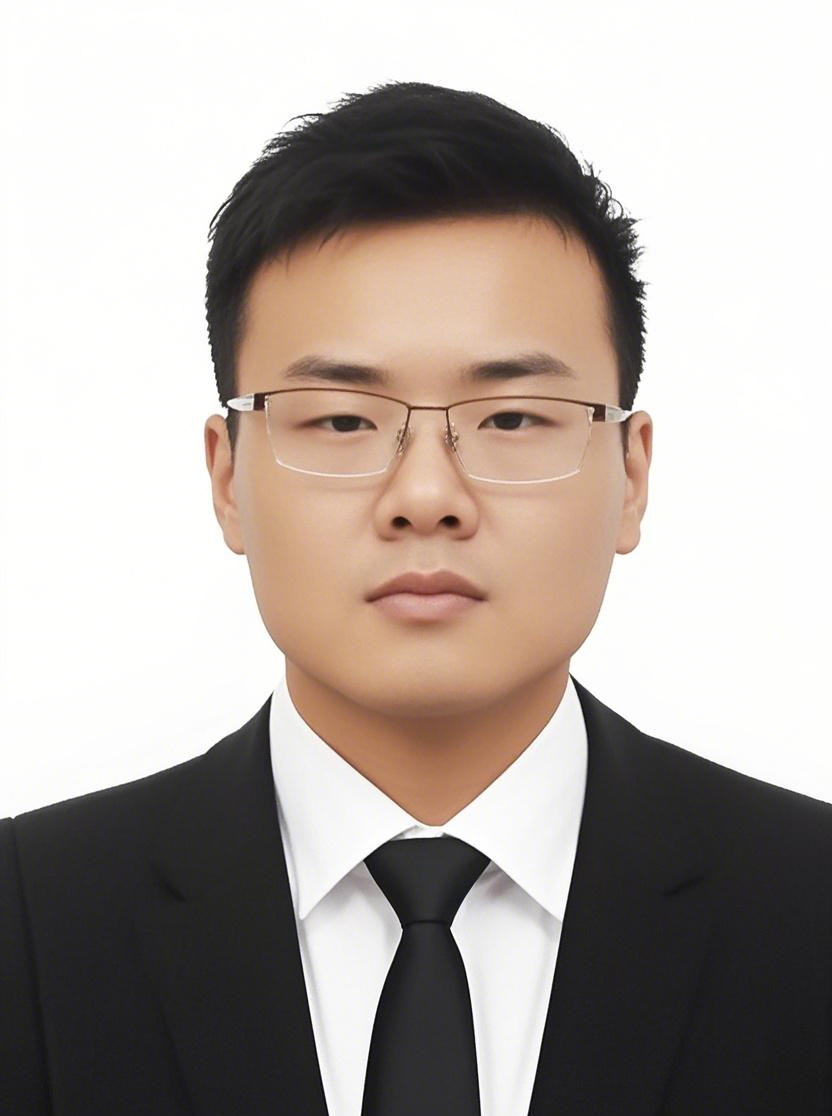}}] {Yu Xie} received the B.S. degree from Suzhou University of Science and Technology, Suzhou, China, in 2022. He is currently working toward the M.S. degree with Jiangnan University. His research interests include digital twin, resource allocation, reinforcement learning, and vehicular edge computing.
	\end{IEEEbiography}
	
	\begin{IEEEbiography}[{\includegraphics[width=1.1in,height=1.4in,clip,keepaspectratio]{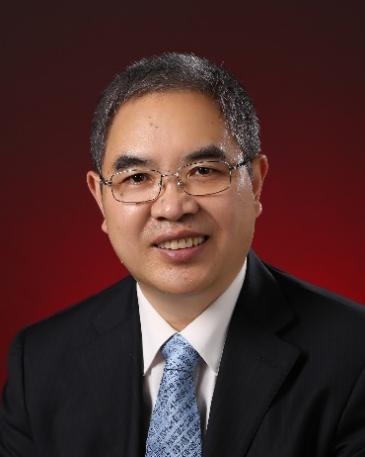}}] {Pingyi Fan} is a professor and the director of open source data recognition innovation center,  Department of Electronic Engineering, Tsinghua University. He is member (Academician) of the united states national academy of artificial intelligence (NAAI), Fellow of IET and IET Fellowship international Assessor. He received Ph.D. degree at the Department of Electronic Engineering of Tsinghua University in 1994. From 1997 to 1999, he visited the Hong Kong University of Science and Technology and the University of Delaware in the United States. He also visited many universities and research institutes in the United States, Europe, Japan, Hong Kong and Singapore. He has obtained many research grants, including national 973 Project, 863 Project, mobile special project and the key R\&D program, national natural funds and international cooperation projects. He has published more than 600 papers (ORCID) including 171 IEEE journals and more than 10 ESI highly cited papers as well as 4 academic books. He also applied for more than 40 national invention patents, 7 international patents. He won 2025 NAAI AI Exploration Award, and 10 best paper awards of IEEE international conferences, including IEEE ICCCS2023 and 2024, ICC2020 and Globecom 2014, and received the best paper award of IEEE TAOS Technical Committee in 2020, the excellent editor award of IEEE TWC (2009), the most popular scholar award 2023 of AEIC, the second natural Prize of CIC (2023) and several international innovation exhibition medals, i.e. Gold Medal at the Russian Invention Exhibition-2024, Silver Medal at Geneva Invention Exhibition-2023, and Silver Medal at Paris Invention Exhibition-2023 etc. and served as the editorial board member of several Journals, including IEEE and MDPI.  He is currently an Associate Editor of IEEE Transactions on Cognitive Communications and Networking (TCCN), the editorial board member of Open Journal of Mathematical Sciences and IAES international journal of artificial intelligence, the deputy director of China Information Theory society, the Co-chair of China's 6G-ANA TG4, and the chairman of Network and Communication Technology Committee of IEEE ChinaSIP. His current research interests are in 6G wireless communication network and machine learning, semantic information theory and generalized information theory, big data processing theory, intelligent network and system detection, etc.
	\end{IEEEbiography}
	
	\begin{IEEEbiography}[{\includegraphics[width=1.1in,height=1.4in,clip,keepaspectratio]{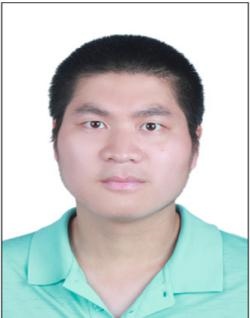}}] {Dong Qin} received the Ph.D. degree in information and communication engineering from Southeast University, Nanjing, China, in 2016. He joined the School of Information Engineering, Nanchang University, in 2016. His current research interests lie in the area of cooperative communication and OFDM techniques.
	\end{IEEEbiography}
	
	\begin{IEEEbiography}[{\includegraphics[width=1.1in,height=1.4in,clip,keepaspectratio]{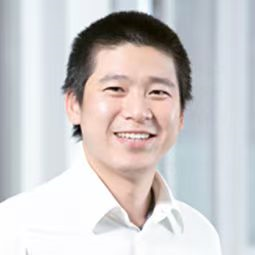}}] {Kezhi Wang} received the Ph.D. degree from the University of Warwick, U.K. He is a Professor with the Department of Computer Science, Brunel University of London, U.K. His research interests include wireless communications, mobile edge computing, and machine learning. He is a Clarivate Highly Cited Researcher in 2023-2024.
	\end{IEEEbiography}
	
	\begin{IEEEbiography}[{\includegraphics[width=1.1in,height=1.4in,clip,keepaspectratio]{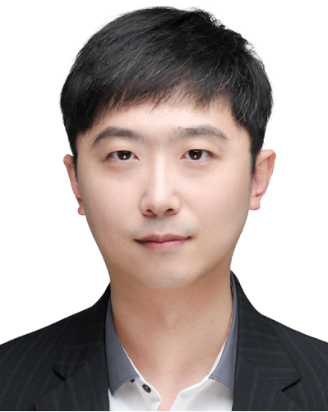}}] {Nan Chen} received the B.E. and M.S. degrees from the College of Electronics and Information Engineering, Tongji University, Shanghai, China, in 2009 and 2012, respectively, and the Ph.D. degree from the Department of Electrical and Computer Engineering, University of Waterloo, Waterloo, ON, Canada, in 2016. He was a Postdoctoral Fellow with the Department of Electrical and Computer Engineering, University of Toronto, Toronto, ON, Canada, from 2017 to 2019. He is currently a Professor with the State Key Laboratory of ISN and with the School of Telecommunications Engineering, Xidian University, Xi’an, Shaanxi, China. He has published over 90 journal papers in IEEE Transactions and other top journals. His current research focuses on B5G/6G, AI-driven future networks, and space–air-ground integrated networks. Prof. Cheng serves as an Associate Editor for IEEE Transactions on Vehicle Technology, IEEE Open Journal of the Communication Society, and Peer-to-Peer Networking and Applications, and serves/served as a guest editor for several journals.
	\end{IEEEbiography}
	
	\begin{IEEEbiography}[{\includegraphics[width=1.1in,height=1.4in,clip,keepaspectratio]{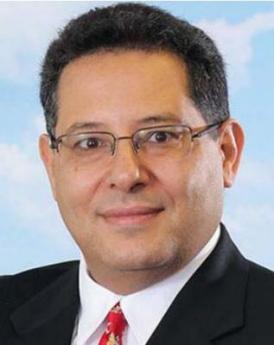}}] {Khaled Ben Letaief} (Fellow, IEEE) received the B.S. (Hons.), M.S., and Ph.D. degrees in electrical engineering from Purdue University, West Lafayette, IN, USA, in December 1984, August 1986, and May 1990, respectively. 
		
	From 1990 to 1993, he was a Faculty Member with The University of Melbourne, Melbourne, VIC, Australia. Since 1993, he has been with The Hong Kong University of Science and Technology (HKUST), Hong Kong, where he is currently the New Bright Professor of Engineering. At HKUST, he has held many administrative positions, including an Acting Provost, the Dean of Engineering, the Head of the Electronic and Computer Engineering Department, and the Director of the Hong Kong Telecom Institute of Information Technology. He is an internationally recognized leader in wireless communications and networks. His research interests include artificial intelligence, mobile cloud and edge computing, tactile Internet, and sixth-genereation (6G) systems. In these areas, he has over 720 articles with over 44,450 citations and an H-index of over 100 along with 15 patents, including 11 U.S. inventions.
		
	Dr. Letaief served as a member for the IEEE Board of Directors from 2022 to 2023. He is a member of the National Academy of Engineering, USA, and the Hong Kong Academy of Engineering Sciences; and a Fellow of the Hong Kong Institution of Engineers. He is well recognized for his dedicated service to professional societies and IEEE, where he served in many leadership positions, including the President of the IEEE Communications Society from 2018 to 2019, the world’s leading organization for communications professionals with headquarter in New York City, and members in 162 countries. He is recognized by Thomson Reuters as an ISI Highly Cited Researcher and was listed among the 2020 top 30 of AI 2000 Internet of Things Most Influential Scholars. He was a recipient of many distinguished awards and honors, including the 2007 IEEE Communications Society Joseph LoCicero Publications Exemplary Award, the 2009 IEEE Marconi Prize Award in Wireless Communications, the 2010 Purdue University Outstanding Electrical and Computer Engineer Award, the 2011 IEEE Communications Society Harold Sobol Award, the 2016 IEEE Marconi Prize Paper Award in Wireless Communications, the 2016 IEEE Signal Processing Society Young Author Best Paper Award, the 2018 IEEE Signal Processing Society Young Author Best Paper Award, the 2019 IEEE Communication Society and Information Theory Society Joint Paper Award, the 2021 IEEE Communications Society Best Survey Paper Award, and the 2022 IEEE Communications Society Edwin Howard Armstrong Achievement Award. He is the Founding Editor-in-Chief of the prestigious IEEE TRANSACTIONS ON WIRELESS COMMUNICATIONS. He has been involved in organizing many flagship international conferences.
	\end{IEEEbiography}


\begin{thebibliography}{11}
		\bibliographystyle{IEEEtran}
		
		\bibitem{ref1}
		B. Cao, Z. Li, X. Liu, Z. Lv and H. He, "Mobility-Aware Multiobjective Task Offloading for Vehicular Edge Computing in Digital Twin Environment," {\it{IEEE Journal on Selected Areas in Communications}}, vol. 41, no. 10, pp. 3046-3055, Oct. 2023.
		
		\bibitem{ref2}
		X. Wang, K. Tao, N. Cheng, Z. Yin, Z Li and Y Zhang, "RadioDiff: An Effective Generative Diffusion Model for Sampling-Free Dynamic Radio Map Construction," {\it{IEEE Transactions on Cognitive Communications and Networking}}, vol. 11, no. 2, pp. 738-750, April. 2025.
		
		\bibitem{ref3}
		R. Sun, N. Cheng, C. Li, F. Chen and W. Chen, "Knowledge-Driven Deep Learning Paradigms for Wireless Network Optimization in 6G," {\it{IEEE Network}}, vol. 38, no. 2, pp. 70-78, March. 2024.
		
		\bibitem{ref4}
		Q. Wu, Y. Zhao and Q. Fan, "Time-Dependent Performance Modeling for Platooning Communications at Intersection," {\it{IEEE Internet of Things Journal}}, vol. 9, no. 19, pp. 18500-18513, Oct. 2022.
		
		\bibitem{ref5}
		Q. Wu, S. Wang, H. Ge, P. Fan, Q. Fan and K. B. Letaief, "Delay-Sensitive Task Offloading in Vehicular Fog Computing-Assisted Platoons," {\it{IEEE Transactions on Network and Service Management}}, pp. 1-1, Oct. 2023.
		
		\bibitem{ref6}
		Q. Wu, X. Wang, Q. Fan, P. Fan, C. Zhang and Z. Li, “High Stable and
		Accurate Vehicle Selection Scheme Based on Federated Edge Learning
		in Vehicular Networks,” {\it{China Communications}}, vol. 20, no. 3, pp. 1–
		17, March. 2023.
		
		\bibitem{ref7}
		Y. Zhang, Y. Zhou, S. Zhang, G. Gui, B. Adebisi, H. Gacanin and H. Sari, "An Efficient Caching and Offloading Resource Allocation Strategy in Vehicular Social Networks," {\it{IEEE Transactions on Vehicular Technology}}, vol. 73, no. 4, pp. 5690-5703, April. 2024.
		
		\bibitem{ref8}
		T. Taleb, K. Samdanis, B. Mada, H. Flinck, S. Dutta and D. Sabella, “On Multi-Access Edge Computing: A Survey of the Emerging 5G Network Edge Cloud Architecture and Orchestration,” {\it{IEEE Communications Surveys Tutorials}}, vol. 19, no. 3, pp. 1657–1681, May. 2017.
		
		\bibitem{ref9}
		Q. Wu, Y. Zhao, Q. Fan, P. Fan, J. Wang and C. Zhang, “MobilityAware Cooperative Caching in Vehicular Edge Computing Based on Asynchronous Federated and Deep Reinforcement Learning,” {\it{IEEE Journal of Selected Topics in Signal Processing}}, pp. 1–16, Jan. 2022.
		
		\bibitem{ref10}
		J. Shen, N. Cheng, X. Wang, F. Lyu, W. Xu and Z. Liu, "RingSFL: An Adaptive Split Federated Learning Towards Taming Client Heterogeneity," {\it{IEEE Transactions on Mobile Computing}}, vol. 23, no. 5, pp. 5462-5478, May. 2024.
		
		\bibitem{ref11}
		N. Cheng, Feng. Lyu, W. Quan, C. Zhou, H. He and W. Shi, "Space/Aerial-Assisted Computing Offloading for IoT Applications: A Learning-Based Approach," {\it{IEEE Journal on Selected Areas in Communications}}, vol. 37, no. 5, pp. 1117-1129, May. 2019.
		
		\bibitem{ref12}
		Y. Dai and Y. Zhang, "Adaptive Digital Twin for Vehicular Edge Computing and Networks," {\it{Journal of Communications and Information Networks}}, vol. 7, no. 1, pp. 48-59, March. 2022.
		
		\bibitem{ref13}
		W. Wang, N Cheng, M. Li, T. Yang, C. Zhou and C. Li, "Value Matters: A Novel Value of Information-Based Resource Scheduling Method for CAVs," {\it{IEEE Transactions on Vehicular Technology}}, vol. 73, no. 6, pp. 8720-8735, June. 2024.
		
		\bibitem{ref14}
		W. Feng, S. Lin, N. Zhang, G. Wang, B. Ai and L. Cai, "Joint C-V2X Based Offloading and Resource Allocation in Multi-Tier Vehicular Edge Computing System," {\it{IEEE Journal on Selected Areas in Communications}}, vol. 41, no. 2, pp. 432-445, Feb. 2023.
		
		\bibitem{ref15}
		Q. Ren, S. Lin, Y. Cai, X. Deng, L. Kong, S. Mumtaz and B. Ai , "Resource Allocation and Slicing Strategy for Multiple Services Co-Existence in Wireless Train Communication Network," {\it{IEEE Transactions on Wireless Communications}}, vol. 24, no. 1, pp. 401-414, Jan. 2025.
		
		\bibitem{ref16}
		Z. Zhang, Q. Wu, P. Fan, N. Cheng, W. Chen and K. B. Letaief, "DRL-Based Optimization for AoI and Energy Consumption in C-V2X Enabled IoV," {\it{IEEE Transactions on Green Communications and Networking}}, doi: 10.1109/TGCN.2025.3531902.
		
		\bibitem{ref17}
		Xueying Gu, Qiong Wu, Pingyi Fan, Nan Cheng, Wen Chen and K. B. Letaief, “DRL-Based Federated Self-Supervised Learning for Task Offloading and Resource Allocation in ISAC-Enabled Vehicle Edge Computing,” {\it{Digital Communications and Networks}}, 2024, doi: https://doi.org/10.1016/j.dcan.2024.12.009.
		
		\bibitem{ref18}
		Xueying Gu, Qiong Wu, Pinyyi Fan, Qiang Fan, Nan Cheng, Wen Chen, and K. B. Letaief, "DRL-Based Resource Allocation for Motion Blur Resistant Federated Self-Supervised Learning in IoV," {\it{IEEE Internet of Things Journal}}, vol. 12, no. 6, pp. 7067-7085, March. 15.
		
		\bibitem{ref19} 
		Yu Xie, Qiong Wu, Pingyi Fan, Nan Cheng, Wen Chen, Jiangzhou Wang, and K. B. Letaief, "Resource Allocation for Twin Maintenance and Task Processing in Vehicular Edge Computing Network," {\it{IEEE Internet of Things Journal}}, doi: 10.1109/JIOT.2025.3576582.
		
		\bibitem{ref20}
		Z. Shao, Q. Wu, P.Fan, N. Cheng, W. Chen, J. Zhou and K. B. Letaief, "Semantic-Aware Spectrum Sharing in Internet of Vehicles Based on Deep Reinforcement Learning," {\it{IEEE Internet of Things Journal}}, vol. 11, no. 23, pp. 38521-38536, Dec. 2024.
		
		
		\bibitem{ref21}
		Z. Zhou, Z. Jia, H. Liao, W. Lu, S. Mutaz, M. Guizani and M. Tariq, "Secure and Latency-Aware Digital Twin Assisted Resource Scheduling for 5G Edge Computing-Empowered Distribution Grids," {\it{IEEE Transactions on Industrial Informatics}}, vol. 18, no. 7, pp. 4933-4943, July. 2022.
		
		\bibitem{ref22}
		C. Zhou, J. Gao, M. Li, N. Cheng, X. Shen and W. Zhuang, "Digital-Twin-Based 3-D Map Management for Edge-Assisted Device Pose Tracking in Mobile AR," {\it{IEEE Internet of Things Journal}}, vol. 11, no. 10, pp. 17812-17826, May. 2024.
		
		\bibitem{ref23}
		Z. Yin, N. Cheng, T. H. Luan, Y. Song and W. Wang, "DT-Assisted Multi-Point Symbiotic Security in Space-Air-Ground Integrated Networks," {\it{IEEE Transactions on Information Forensics and Security}}, vol. 18, pp. 5721-5734, September. 2023.
		
		\bibitem{ref24}
		Z. Wang, G. Rohit, K. Han, H. Wang, G. Akila, A. Nejib and T. Prashant, "Mobility Digital Twin: Concept, Architecture, Case Study, and Future Challenges," {\it{IEEE Internet of Things Journal}}, vol. 9, no. 18, pp. 17452-17467, Sept. 2022.
		
		\bibitem{ref25}
		J. Zheng, T. H. Luan, Y. Hui, Z. Yin, N. Chen, L. Gao and L. Cai, "Digital Twin Empowered Heterogeneous Network Selection in Vehicular Networks With Knowledge Transfer," {\it{IEEE Transactions on Vehicular Technology}}, vol. 71, no. 11, pp. 12154-12168, Nov. 2022.
		
		\bibitem{ref26}
		C. Xu, Z. Tang, H. Yu, P. Zeng and L. Kong, "Digital Twin-Driven Collaborative Scheduling for Heterogeneous Task and Edge-End Resource via Multi-Agent Deep Reinforcement Learning," {\it{IEEE Journal on Selected Areas in Communications}}, vol. 41, no. 10, pp. 3056-3069, Oct. 2023.
		
		\bibitem{ref27}
		T. Liu, L. Tang, W. Wang, Q. Chen and X. Zeng, "Digital-Twin-Assisted Task Offloading Based on Edge Collaboration in the Digital Twin Edge Network,"  {\it{IEEE Internet of Things Journal}}, vol. 9, no. 2, pp. 1427-1444, 15 Jan. 2022.
		
		\bibitem{ref28}
		E. Zhang, L. Zhao, N. Lin, W. Zhang, A. Hawbani and G. Min, "Cooperative Task Offloading in Cybertwin-Assisted Vehicular Edge Computing,"  {\it{Proc. IEEE 20th Int. Conf. Embedded Ubiquitous Comput. (EUC)}},
		Dec. 2022, pp. 66–73.
		
		\bibitem{ref29}
		K. Zhang, J. Cao and Y. Zhang, "Adaptive Digital Twin and Multiagent Deep Reinforcement Learning for Vehicular Edge Computing and Networks," {\it{IEEE Transactions on Industrial Informatics}}, vol. 18, no. 2, pp. 1405-1413, Feb. 2022.
		
		\bibitem{ref30}
		X. Liao, X. Zhao, Z. Wang, Z. Zhao, K. Han, R. Gupta, M. J. Barth and G. Wu, "Driver Digital Twin for Online Prediction of Personalized Lane-Change Behavior," {\it{IEEE Internet of Things Journal}}, vol. 10, no. 15, pp. 13235-13246, 1 Aug. 2023.
		
		\bibitem{ref31}
		K. Zhang, J. Cao, S. Maharjan and Y. Zhang, "Digital Twin Empowered Content Caching in Social-Aware Vehicular Edge Networks," {\it{IEEE Transactions on Computational Social Systems}}, vol. 9, no. 1, pp. 239-251, Feb. 2022.
		
		\bibitem{ref32}
		J. Zheng, T. H. Luan, Y. Zhang, R. Li, Y. Hui, L. Gao and M. Dong, "Data Synchronization in Vehicular Digital Twin Network: A Game Theoretic Approach," {\it{IEEE Transactions on Wireless Communications}}, vol. 22, no. 11, pp. 7635-7647, Nov. 2023.
		
		\bibitem{ref33}
		L. Zhao, Z. Zhao, E. Zhang, A. Hawbani, A. Y. Al-Dubai, Z. Tan and A. Hussian, "A Digital Twin-Assisted Intelligent Partial Offloading Approach for Vehicular Edge Computing," {\it{IEEE Journal on Selected Areas in Communications}}, vol. 41, no. 11, pp. 3386-3400, Nov. 2023.
		
		\bibitem{ref34}
		B. Li, W. Xie, Y. Ye, L. Liu and Z. Fei, "FlexEdge: Digital Twin-Enabled Task Offloading for UAV-Aided Vehicular Edge Computing," {\it{IEEE Transactions on Vehicular Technology}}, vol. 72, no. 8, pp. 11086-11091, Aug. 2023.
		
		\bibitem{ref35}
		H. Zhou, C. Hu, D. Yuan, Y. Yuan, D. Wu, X. Liu and C. Zhang, "Large Language Model (LLM)-enabled In-context Learning for Wireless Network Optimization: A Case Study of Power Control", {\it{arXiv preprint arXiv:2408, 00214}}, 2024.
		
		\bibitem{ref36}
		H.Zhou, C. Hu, D. Yuan, Y. Yuan, D. Wu, X. Liu, Z. Han and C. Zhang, "Generative AI as a Service in 6G Edge-Cloud: Generation Task Offloading by In-context Learning", {\it{arXiv preprint arXiv:2408, 02549}}, 2024.
		
		\bibitem{ref37}
		W. Lee and J. Park, "LLM-Empowered Resource Allocation in Wireless Communications Systems", {\it{arXiv preprint arXiv:2408, 02944}}, 2024.
		
		\bibitem{ref38}
		M. Fu, P. Wang, M. Liu, Z. Zhang and X. Zhou, "IoV-BERT-IDS: Hybrid Network Intrusion Detection System in IoV Using Large Language Models," {\it{IEEE Transactions on Vehicular Technology}}, vol. 74, no. 2, pp. 1909-1921, Feb. 2025.
		
		\bibitem{ref39}
		Q. Liu, J. Mu, D. Chen, R. Zhang, Y. Liu and T. Hong, "LLM Enhanced Reconfigurable Intelligent Surface for Energy-Efficient and Reliable 6G IoV," {\it{IEEE Transactions on Vehicular Technology}}, vol. 74, no. 2, pp. 1830-1838, Feb. 2025.
		
		\bibitem{ref40}
		X. Chen, X. Lu, Q. Li, D. Li and F. Zhu, "Integration of LLM and Human–AI Coordination for Power Dispatching With Connected Electric Vehicles Under SAGVNs," {\it{IEEE Transactions on Vehicular Technology}}, vol. 74, no. 2, pp. 1992-2002, Feb. 2025.
		
		\bibitem{ref41}
		Q. Wu, W. Wang, P. Fan, Q. Fan, J. Wang and K. B. Letaief, "URLLC-Awared Resource Allocation for Heterogeneous Vehicular Edge Computing," {\it{IEEE Transactions on Vehicular Technology}}, vol. 73, no. 8, pp. 11789-11805, Aug. 2024.
		
		\bibitem{ref42}
		X. Tan, M. Wang, T. Wang, Q. Zheng, J. Wu and J. Yang, "Adaptive Task Scheduling in Digital Twin Empowered Cloud-Native Vehicular Networks," {\it{IEEE Transactions on Vehicular Technology}}, vol. 73, no. 6, pp. 8973-8987, June 2024.
		
		\bibitem{ref43}
		M. J. Neely, “Stochastic Network Optimization with Application to Communication and Queueing Systems,” {\it{Synthesis Lectures on Com munication Networks}}, vol. 3, no. 1, pp. 1–211, Sept. 2010.
		
		\bibitem{ref44}
		L. Tang, Z. Cheng, J. Dai, H. Zhang and Q. Chen, "Joint Optimization of Vehicular Sensing and Vehicle Digital Twins Deployment for DT-Assisted IoVs," {\it{IEEE Transactions on Vehicular Technology}}, vol. 73, no. 8, pp. 11834-11847, Aug. 2024.
		
		\bibitem{ref45}
		 Y. Li, S. Xia, M. Zheng, B. Cao, and Q. Liu, “Lyapunov Optimization Based Trade-Off Policy for Mobile Cloud Offloading in Heterogeneous Wireless Networks,” {\it{IEEE Transactions on Cloud Computing}}, vol. 10, no. 1, pp. 491–505, Jan.-March 2022.
\bibitem{106}
Q.~Wu and J.~Zheng, ``Performance modeling of ieee 802.11 dcf based fair
  channel access for vehicular-to-roadside communication in a non-saturated
  state,'' in \emph{2014 IEEE International Conference on Communications
  (ICC)}.\hskip 1em plus 0.5em minus 0.4em\relax IEEE, 2014, pp. 2575--2580.		
  \bibitem{107}
K.~Qi, Q.~Wu, P.~Fan, N.~Cheng, W.~Chen, J.~Wang, and K.~B. Letaief,
  ``Deep-reinforcement-learning-based aoi-aware resource allocation for
  ris-aided iov networks,'' \emph{IEEE Transactions on Vehicular Technology},
  2024.		
\bibitem{108}
Y.~Dong, Z.~Chen, S.~Liu, P.~Fan, and K.~B. Letaief, ``Age-upon-decisions
  minimizing scheduling in internet of things: To be random or to be
  deterministic?'' \emph{IEEE Internet of Things Journal}, vol.~7, no.~2, pp.
  1081--1097, 2019.
  \bibitem{109}
T.~Li, P.~Fan, Z.~Chen, and K.~B. Letaief, ``Optimum transmission policies for
  energy harvesting sensor networks powered by a mobile control center,''
  \emph{IEEE Transactions on Wireless Communications}, vol.~15, no.~9, pp.
  6132--6145, 2016.
  
  \bibitem{110}
Y.~Yang and P.~Fan, ``Doppler frequency offset estimation and diversity
  reception scheme of high-speed railway with multiple antennas on separated
  carriage,'' \emph{Journal of Modern Transportation}, vol.~20, no.~4, pp.
  227--233, 2012.
  
\bibitem{111}
P.~Fan, C.~Feng, Y.~Wang, and N.~Ge, ``Investigation of the time-offset-based
  qos support with optical burst switching in wdm networks,'' in \emph{2002
  IEEE International Conference on Communications. Conference Proceedings. ICC
  2002 (Cat. No. 02CH37333)}, vol.~5.\hskip 1em plus 0.5em minus 0.4em\relax
  IEEE, 2002, pp. 2682--2686.		
	\end{thebibliography}
\end{document}